\documentclass[a4paper,amsmath,amssymb,12pt]{revtex4}
\usepackage{url}
\usepackage{amssymb}
\usepackage{graphicx}
\usepackage{epsfig}

\usepackage{float}
\usepackage{bm}
\usepackage[font=small,labelfont=bf,labelsep=space]{caption}
\usepackage{refstyle}

\begin{document}
%
\title{Low-Dimensional Few-Body Processes in Confined Geometry of Atomic and Hybrid
                                                           Atom-Ion Traps}
%
%
%
%
%
\author{Vladimir S. Melezhik} 
\affiliation{Bogoliubov Laboratory of Theoretical Physics, Joint Institute for Nuclear Research, Dubna, Moscow Region 141980, Russian Federation}
\affiliation{Peoples' Friendship University of Russia (RUDN University) Miklukho-Maklaya st. 6, Moscow,  117198, Russian Federation}

\begin{abstract}
We have developed an efficient approach for treating low-dimensional few-body processes in confined geometry of atomic and hybrid atom-ion traps. It based on the split-operator method in 2D discrete-variable representation (DVR) suggested by V. Melezhik for integration of the few-dimensional time-dependent Schr\"odinger equation. We give a brief review of the application to resonant ultracold atomic processes and discuss our latest results on hybrid atomic-ion systems. Prospects for the application of the method in other hot problems of the physics of low-dimensional few-particle systems are also discussed.
\end{abstract}
\maketitle              

\section{Introduction}
Impressive progress of the physics of ultracold
quantum gases has stimulated the necessity of detailed and
comprehensive investigations of collisional processes in the
confined geometry of atomic and ionic traps. The traditional
free-space scattering theory is no longer valid here and the
development of the low-dimensional few-body theory including the influence
of the confinement is needed. In our works we have developed
quantitative models [1-4] for pair collisions in tight atomic
waveguides and have found several novel effects in its
application: the confinement-induced resonances (CIRs) in
multimode regimes including effects of transverse excitations and
deexcitations [2], the so-called dual CIR yielding a complete
suppression of quantum scattering [1], and resonant molecule
formation with a transferred energy to center-of-mass excitation
while forming molecules [5]. Last effect was confirmed experimentally in [6]. Our calculations have also been used
for planning and interpretation of the Innsbruck experiment where
CIRs in ultracold Cs gas were observed [7]. Mention also the
calculation of the Feshbach resonance shifts and widths induced by
atomic waveguides [8]. In the frame of our
approach we have predicted dipolar CIRs [9] which may pave the way for the experimental realization
of, e.g., Tonks-Girardeau-like or super-Tonks-Girardeau-like
phases in effective one-dimensional dipolar gases.

Our latest results on hybrid atomic-ion systems and prospects are discussed in this report.

\section{Atom-Ion Collisions in Hybrid Atom-Ion Traps}

Recently, we have predicted the atom-ion CIRs [10] which are important for a hot problem of control of
the confined hybrid atom-ion systems having many promising applications [11]. The condition of appearance of CIR in a atom-ion collision confined in a harmonic waveguide-like trap was found in [10] in "static" ion approximation. This approach, when one neglects by the ion motion, is well defined for for the Li-Yb$^+$ collision considered in [10]. However, in real experiments an actual problem is controlling of the unremovable effect of ion micromotion in the ion Paul traps [11].

\begin{figure}[h]
\vspace{-.5cm}
\includegraphics[width=1.1\linewidth]{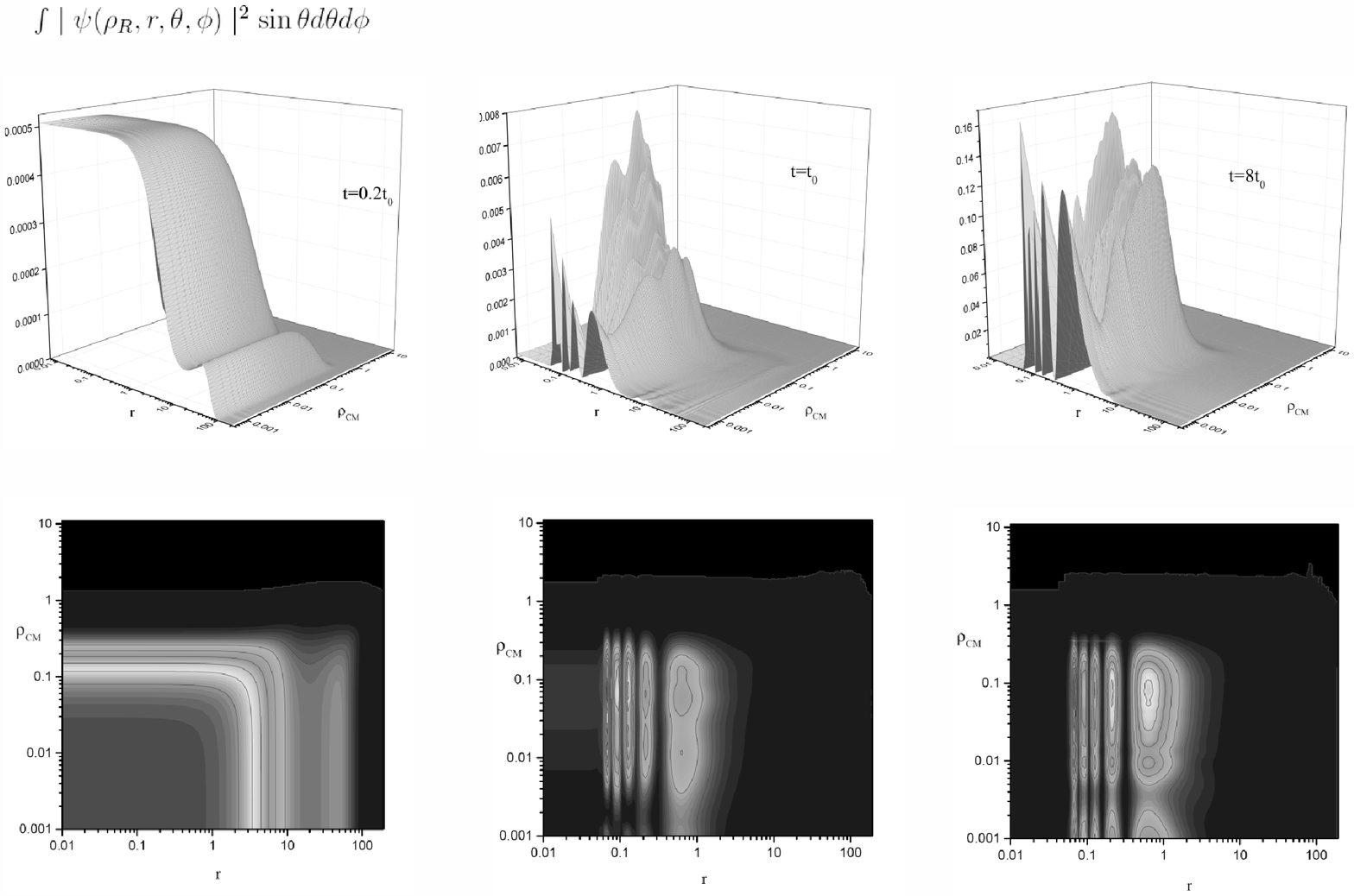}
\vspace{-0.5cm}
\caption{Evolution in time of the atom-ion probability density distribution $W(r,\rho_R,t)=\int |\psi(r,\rho_R,t)|^2\sin\theta d\theta\phi$ in the process of the confined collision of the $^6$Li atoms with $^{171}$Yb$^+$ ions. Top row of graphs demonstrates the probability density $W(r,\rho_R,t)$ calculated at three time points during the collision and the bottom row is the corresponding contour plots of these probabilities $W(r,\rho_R,t)$. The time unit is $t_0=2\pi/\omega_A$.}
\end{figure}
To evaluate the effect of the ion motion on the CIR we performed full quantum calculation for the $^6$Li-atom scattering by $^{171}$Yb$^+$ for a special case of harmonic transversal traps with $\omega_A\ll\omega_I$ frequencies for atom (A) and ion (I)
$$
U(\rho_A,\rho_I)=\frac{1}{2} (m_A \omega_A^2 \rho_A^2 + m_I \omega_I^2 \rho_I^2)
$$
(where $\rho_i = r_i sin \theta_i$). For that, we have integrated the 4D time-dependent Schr\"odinger equation
with the Hamiltonian
\begin{equation}
H(\rho_R,{\bf r})= H_{CM}(\rho_R) + H_{rel} ({\bf r}) +
W(\rho_R,{\bf r}) \,\,.
\end{equation}
by using computational scheme developed earlier for confined distinguishable atom collisions [1,3,5].
Here ($\hbar =1$)
\begin{equation}
H_{CM}= -\frac{1}{2M}(\frac{\partial
^2}{\partial \rho_{R}^{2}} +\frac{1}{\rho_{R}^{2}}
\frac{\partial^2}{\partial \phi^2} +\frac{1}{4\rho_{R}^{2}})
+\frac{1}{2}(m_{A}\omega_{A}^{2}+m_{I}\omega_{I}^{2})\rho_{R}^{2}
\end{equation}
and
\begin{equation}
H_{rel} = -\frac{1}{2\mu}\frac{\partial ^{2}}{\partial r^{2}}+
\frac{L^{2}(\theta,\phi)}{2\mu
r^{2}}+\frac{\mu^2}{2}(\frac{\omega_{A}^{2}}{m_{A}}+\frac{\omega_{I}^{2}}{m_{I}})
\rho^2+V_{AI}(r)
\end{equation}
describe the CM and relative (rel) atom-ion motions. The potential  $V_{AI}(r)$ describes the atom-ion interaction,
$\rho_R$ and $\bf{r} = \bf{r}_A - \bf{r}_I\mapsto (r, \theta, \phi) \mapsto (\rho, \phi, z)$ are the polar
radial CM and the relative coordinates and
$M=m_A+m_I$, $\mu = m_A m_I/ M$. The term $\frac{L^2(\theta,\phi)}{2\mu r^2}$
represents the angular part of the kinetic energy operator of the relative
atom-ion motion.
The term
\begin{equation}
W(\rho_R,{\bf r})=\mu(\omega_{A}^2-\omega_{I}^{2}) r \rho_{R} \sin \theta \cos \phi
\end{equation}
leads to a coupling of the CM and relative
motion, i.e. to the nonseparability of the quantum two-body problem in confined geometry of the
harmonic trap.

In Fig.1 we present the calculated time-evolution of the probability density distribution of $^6$Li and $^{171}$Yb$^+$ near the CIR in the harmonic waveguides.
This quantum calculation confirms the surviving of the CIR in the case of ion-motion and demonstrates the molecule ion LiYb$^+$ formation during this collision.

\section{Conclusion}
The efficiency of the splitting-up method based on the 2D DVR for the time-dependent
Schr\"odinger equation makes the method promising in application
to actual problems of low-dimensional few-body physics in atomic and atom-ion traps. One can mention  the problem of ultracold atomic
collisions in anharmonic and asymmetric waveguides and in quasi-2D confining traps.
Of great interest in connection with possible important applications is the two-center ( and N-center) problem in a confining trap \cite{12,13}. Note also a collisional three-body problem in tight traps,
and non-linear time-dependent Schr\"odinger equation with a few
spatial variables arising in physics of  Bose-Einstein
condensates.

This work was supported by the Russian Foundation for Basic Research, Grant No. 18-02-00673 and the ``RUDN University Program 5-100''.

%
%

\end{document}